\documentclass{aa}
\usepackage{graphicx}
\usepackage[figuresright]{rotating}
\usepackage{longtable,lscape}

\begin{document}

\title{The VVDS-VLA Deep Field:}
\subtitle{III. GMRT observations at 610 MHz and the radio spectral index 
properties of the sub-mJy population\thanks{Table 1 is only available in
electronic form at the CDS via http://cdsweb.u-strasbg.fr/cgi-bin/qcat?J/A+A}}
\author{M. Bondi\inst{1} \and P. Ciliegi\inst{2} \and T. Venturi\inst{1}
\and D. Dallacasa\inst{3,1} \and S. Bardelli\inst{2} \and E. Zucca\inst{2}
\and R.M. Athreya\inst{4} \and L. Gregorini\inst{5,1} \and A. Zanichelli\inst{1}
\and O. Le F\`evre\inst{6} \and T. Contini\inst{7}
\and B. Garilli\inst{8} \and A. Iovino\inst{9} \and S. Temporin\inst{9}
\and D. Vergani\inst{8}
}

\institute{INAF - Istituto di Radioastronomia, 
Via Gobetti 101, I-40129 Bologna, Italy 
\and
INAF - Osservatorio Astronomico di Bologna, Via Ranzani 1, 
I-40127, Bologna, Italy 
\and
Universit\`a degli Studi di Bologna, 
Dipartimento di Astronomia, Via Ranzani 1, I-40127 Bologna, Italy
\and
NCRA, Tata Institute of Fundamental Research Pune University Campus, Post
Bag 3, Ganeshkhind Pune 411007, India
\and
Universit\`a degli Studi di Bologna, 
Dipartimento di Fisica, Via Irnerio 46, I-40126 Bologna, Italy
\and
Laboratoire d'Astrophysique de Marseille, UMR 6110 CNRS-Universit\'e de
Provence,  BP8, 13376 Marseille Cedex 12, France
\and
Laboratoire d'Astrophysique de l'Observatoire Midi-Pyr\'en\'ees (UMR
5572) -
14, avenue E. Belin, F31400 Toulouse, France
\and
IASF-INAF - via Bassini 15, I-20133, Milano, Italy
\and
INAF-Osservatorio Astronomico di Brera - Via Brera 28, Milan,
Italy
}
%\date{}
\date{Received 22 September 2006 / Accepted 17 October 2006}
\abstract
{}
{We present the low frequency (610 MHz) radio source counts of the
VVDS-VLA field and investigate the radio spectral index properties
of the sub-mJy population.}
{We use new deep (r.m.s.$\simeq 50$ $\mu$Jy/beam) observations of the VVDS-VLA 
field obtained at 610 MHz with the GMRT and matched in resolution (6 arcsec) 
with already available VLA data at 1.4 GHz on the same field.}
{We find evidence of a change of the dominant population of radio sources 
below 0.5 mJy (at 1.4 GHz): between 0.15 and 0.5 mJy the median spectral index
is significantly flatter ($\alpha=-0.46\pm 0.03$) than that of brighter sources
($\alpha=-0.67\pm 0.05$). A relevant contribution below 0.5 mJy from a 
population of flat spectrum low luminosity compact AGNs and radio quiet QSOs 
could explain this effect. At even fainter flux density, between
0.10 and 0.15 mJy at 1.4 GHz, the median spectral index steepens again
($\alpha=-0.61\pm 0.04$) suggesting that the contribution of starburst galaxies
becomes important below $\sim 0.2$ mJy. Finally we present a sample of
58 candidate ultra-steep sources with radio flux density from one to two
orders of magnitude lower than any other sample of such objects.}
{}
 

\maketitle
\section{Introduction}

The "normal" population of radio sources with flux densities greater 
than a few mJy is fully explained in terms of radio loud active galactic 
nuclei (AGN) hosted by elliptical galaxies. 
At fainter flux densities, now extending to $\mu$Jy levels, the 
radio source counts are increasingly dominated by contributions from 
other populations, such as radio-quiet AGNs mainly hosted by elliptical
galaxies and starbursts in late-type galaxies, whose radio emission 
is the result of massive star formation and associated supernovae activity 
(\cite{Wind85}; 
\cite{Benn93}; \cite{Norr05}; \cite{Hamm95}; 
\cite{Rich98}; \cite{Afon05}; Pran\-do\-ni et al. 2001\nocite{Pran01}; 
\cite{Geor99}; \cite{Grup99}).

Far-infrared and X-ray observations have also provided further support for
both starburst and AGN processes in the sub-mJy radio sources
(\cite{Afon01}, 2006\nocite{Afon06}; \cite{Geor03},~2004\nocite{Geor04})
In the radio domain, the properties of the faint 
radio population have been mainly obtained from samples selected at 1.4 GHz 
(e.g., \cite{Rich00}; \cite{Cili99}; \cite{Bond03}; \cite{Hopk03}; 
\cite{Morg04}; \cite{Huyn05}). Subsequent follow-up observations, carried out 
at higher frequencies (5 GHz or 8.4 GHz; e.g., \cite{Donn87}; \cite{Cili03};
\cite{Foma06}) have been used to infer the triggering mechanism of the 
radio emission. 
In fact, the radio/optical morphology (compact or extended) and the radio 
spectrum (flat or steep) can help discriminate between emission 
from starbursts and/or AGNs and understanding the interaction between them. 
However, the angular size of the radio emission can be a definite 
discriminator only at very high resolution ($<0.05$ arcsec, \cite{Garr01}; 
\cite{GWM05}), while the radio spectral 
properties, despite the extensive work done in recent years, can be
interpreted in different ways and suffer from small number statistics.   

Analyzing rather small samples ($\la 60$ sources) of sub-mJy sources, 
several authors (\cite{Donn87}; \cite{Grup97}; \cite{Cili03}) have found 
that the fraction of flat spectrum objects increases going to fainter density 
flux. 
Recently, with a more robust statistical significance, Prandoni et al. 
(2006)\nocite{Pran06}
confirmed this result for a sample of 131 radio sources extracted from the 
ATESP survey. However, it is not yet established at which flux density level 
the flattening of the spectral index occurs.
 
At $\mu$Jy level, Windhorst et al. (1993)\nocite{Wind93} found a median 
spectral index of $-0.35$ ($S\propto \nu^\alpha$), which, combined with the
rather large linear size of the 
radio sources, has suggested synchrotron emission in distant galactic disks 
for the extended steep-spectrum sources and thermal Bremsstrahlung 
from large-scale star formation for the extended flat-specrtum 
sources. Recently, from the analysis of a sample of 47 radio sources, 
Fomalont et al. (2006)\nocite{Foma06} found that the spectral index may 
steepen for sources fainter than 75 $\mu$Jy and that in at most 40\% of the 
$\mu$Jy population the radio emission is associated with AGN emission, while  
the rest is mostly a consequence of star formation.      
  
In this context the radio data of the VVDS-VLA Deep survey at 1.4 GHz 
(\cite{Bond03}, Paper I) are perfectly suited for the study the properties 
of the sub-mJy population. 
From the deep radio observations carried out with the VLA, a catalogue 
containing 1054 radio sources detected down to a 5$\sigma$ limit of $\simeq$ 
80$\mu$Jy was 
extracted. About 74$\%$ of the whole sample was optically identified using 
UBVRI and K data at an equivalent limiting magnitude of $V_{AB}=26.2$
(\cite{Cili05}, Paper II).
The color and photometric redshift analysis of the optical counterparts 
show that radio detection preferentially selects galaxies with 
higher intrinsic optical luminosity and that most of the faintest radio 
sources are likely to be associated with 
relatively low radio luminosity objects at relatively modest redshift, 
rather than radio-powerful, AGN type objects at high redshift. 
The authors found that the majority of radio sources below ~0.15 mJy are 
indeed late-type star-forming galaxies and that the radio sources without 
an optical counterpart probably contain a significant fraction of 
obscured and/or high redshift galaxies.

\begin{figure*}
\includegraphics[width=12cm,angle=-90]{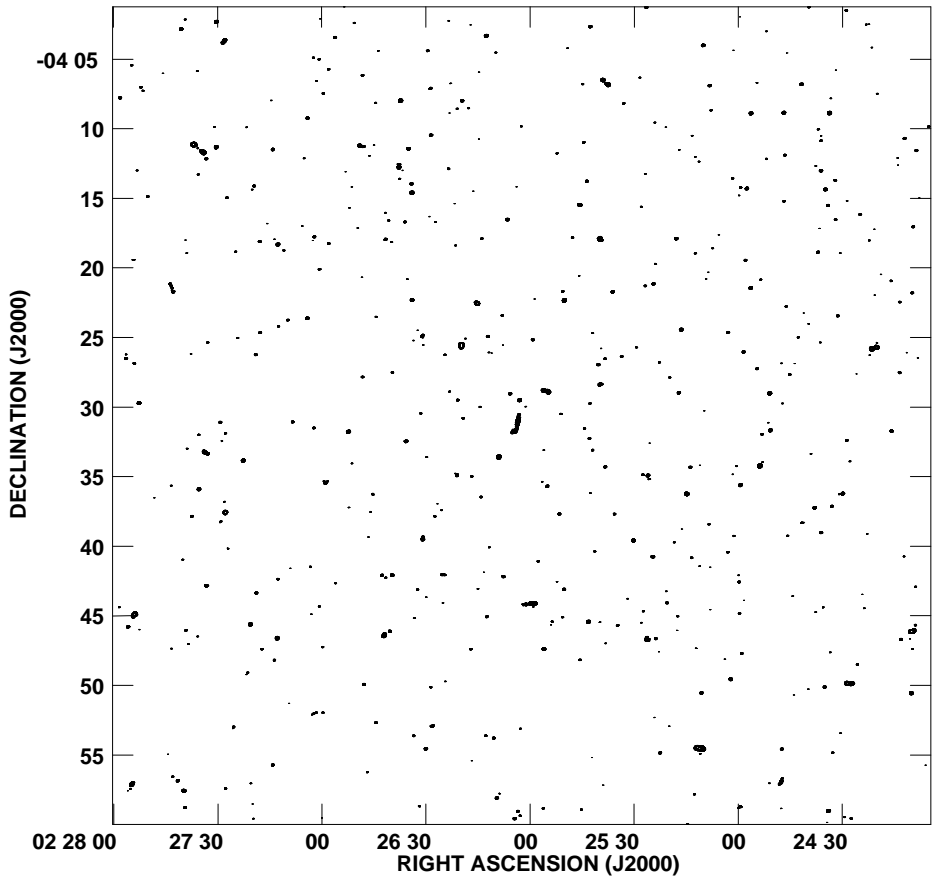}
\caption[]{Radio mosaic at 610 MHz of the F02 field with a resolution of 6
arcsec.
Contours are in units of signal-to-noise ratio, and the first contour is 5 
times the
local r.m.s. The average r.m.s on the whole image is $50\mu$Jy/beam.}
\label{field_radio}
\end{figure*}

In this paper we present GMRT radio observations at 610 MHz of the
VVDS-VLA field (Sect. 2). We discuss the methods used to derive the 
$5\sigma$ catalogue of 514 radio sources (Sect. 3) and present the 
radio source counts at 610 MHz down to 0.3 mJy (Sect. 4). 
Furthermore, using the 1.4 GHz catalogue obtained in Paper I, we
investigate the radio spectral index properties of the sub-mJy radio sources
and
the relationships between the radio spectral index and the optical properties
(Sect. 5). 
Finally we introduce a small sample of faint candidate ultra-steep spectrum 
radio sources (Sect. 6).

\section{GMRT observations \& data reduction}

\subsection{Observations}

The VVDS-VLA field (RA(J2000) $=$ 02:26:00 DEC(J2000) $=$ $-$04:30:00, 
hereafter F02 field) 
was observed at the radio frequency of 610 MHz with 
the Giant Meter-Wave Radio Telescope (GMRT) in August 2004 for a total time 
of 24 
hours (plus a short backup observation of about 4 hours in December 2004 to 
recover time lost during the run). 
At 610 MHz the GMRT has an angular resolution of about 6 arcsec and
a primary beam of 43 arcmin (FWHM).
To observe the whole field with a sensitivity as uniform as possible,
the 1 sq. deg. area of F02  was covered with a grid of 5 pointings:
one in the field center and the other four displaced by 15 arcmin in both
right ascension and declination from the central pointing. 
%The position of the pointing centers is shown in Fig.~\ref{noise_map}.
Each pointing was observed for about 5.5 hours.

Both the lower and upper sidebands, centered around 610 MHz, were recorded 
yielding to a total bandwidth of 32 MHz. 
The observations were carried out using 128 channels
in each band and a spectral resolution of 125 kHz.
In this way it was also possible to reduce the effects of narrow-band
interferences since only the channels affected by the interferences, 
instead of the whole bandwidth, can be removed from the data.
The radio sources 3C48 and 3C147 were observed at the beginning and at the
end of the run as primary flux density and bandpass calibrators.
The nearby source J0241$-$082 was observed every 30 minutes to provide 
secondary amplitude and phase calibration.

\subsection{Data reduction}

The data were reduced and analyzed using the package AIPS developed by the
National Radio Astronomy Observatory.
The amplitude and bandpass calibration were derived from daily observations of 
3C~48 and 3C~147 assuming a flux density of 29.43 Jy and 38.26 Jy,
respectively.
After the bandpass calibration the central channels of each sideband were
averaged to 6 channels of 1.75 MHz each, producing a total effective
bandwidth of 21 MHz.
The AIPS tasks UVLIN and CLIP were used to flag bad visibility data
resulting from residual radio frequency interferences, receiver problems or 
correlator failures. 

For each of the five pointings we imaged a field of about $51\times 51$
arcmin ($2048\times 2048$ pixels with a pixel size of 1.5 arcsec),
roughly the area of the whole primary beam,
along with a number of smaller images centered on off-axis sources that can 
produce confusing grating rings in the imaged area. 
To avoid distortions due to the use of two-dimensional FFT to approximate
the curved celestial sphere, the $2048\times 2048$ pixels area of each pointing
was not deconvolved as a single image, but was split into  four $1024\times 
1024$ sub-images (e.g., \cite{Perl99}).

At the end of the self-calibration deconvolution iteration scheme, the
sub-images of all the 5 fields were combined together using the AIPS task 
FLATN and corrected for the primary beam response of GMRT at 610 MHz
to produce the final mosaic with a resolution of 6 arcsec.
The average noise over the whole 1 square degree field in the mosaic map
is about 50 $\mu$Jy/beam providing a dynamic range greater than 2000,
consistent with that expected.
In Fig.~\ref{field_radio} we show the 
final image of the 1 square degree field in units of signal-to-noise ratio.

\section{The 610 MHz radio source catalogue}

\subsection{The noise map} 

\begin{figure}
\begin{center}
\includegraphics[width=8cm,angle=-90]{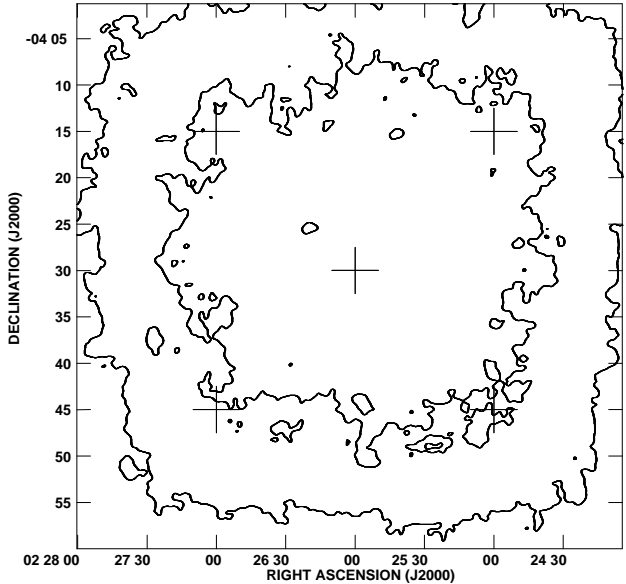}
\caption[]{A contour plot of the noise map obtained with SExtractor. 
The inner contour contains the region of the image with noise less than 50
$\mu$Jy/beam, the outer contour the region with noise less than 70
$\mu$Jy/beam. Crosses show the centers of 5 individual pointings.
}
\label{noise_map}
\end{center}
\end{figure}

To select a sample of sources above a given threshold, defined in 
terms of local signal-to-noise ratio, we derived the noise image following
the same procedure adopted for the 1.4 GHz observations and using the 
software package SExtractor (\cite{BA96}).
SExtractor computes 
an estimator for the local background in each mesh of a grid that covers the
whole image (see \cite{BA96} for more details). To be fully consistent with
the analysis of the 1.4 GHz data we adopted the same mesh size,  20 pixels,
corresponding to 30 arcsec. 
Figure~\ref{noise_map} shows a contour plot of the 
noise image.
The noise is reasonably uniform in the inner $40\times 40$ arcmin and starts
increasing closer to the image borders.
The pixel value distribution has a peak at about 50 $\mu$Jy/beam, well in 
agreement with the average noise value reported in the previous section.

\subsection{The source detections} 

We selected the 1 square degree region centered on the F02 field
and 
within this region we extracted all the radio components with a peak flux 
S$_{\rm P}>$150 $\mu$Jy/beam ($\sim 3\sigma$), using the AIPS task SAD.
For each selected component, the peak brightness and total flux density, 
the position, and the 
size are estimated using a Gaussian fit. 
However, for faint components the Gaussian 
fit may be unreliable and a better estimate of the peak flux 
S$_{\rm P}$ (very important since it is used for the selection) and of the 
component position is obtained with a non-parametric interpolation of the 
pixel values around the fitted position. 
Therefore, for all the components found by SAD, we derived the  
peak flux S$_{\rm P}$ and the position using a 
second-degree interpolation fit with the task MAXFIT.
Only the components with a signal-to-noise ratio (derived as the ratio 
between the MAXFIT peak flux and the local noise from the noise map)
greater than or equal to 5 have been included in the sample.  
Around the brightest sources ($> 10$ mJy/beam), residual sidelobes can be 
mistakenly identified with real components by SAD.
For this reason we visually inspected the region around bright
sources to remove sidelobe spikes.

In this way, we selected a sample of 557 components above the local $5\sigma$
threshold. All the components have  flux densities greater than 0.2 mJy at 
610 MHz.
The 557 components correspond to 514 sources, since 17 sources are 
considered as multiple, i.e., fitted with at least two separate
components, following the same criteria adopted for the 1.4 GHz catalogue.
All the 17 multiple sources at 610 MHz were also found to be multiple 
at 1.4 GHz.

\subsection{Resolved and unresolved sources} 

To discriminate between resolved and unresolved sources we used the ratio
between total and peak flux density.
\begin{figure}
\includegraphics[width=8cm]{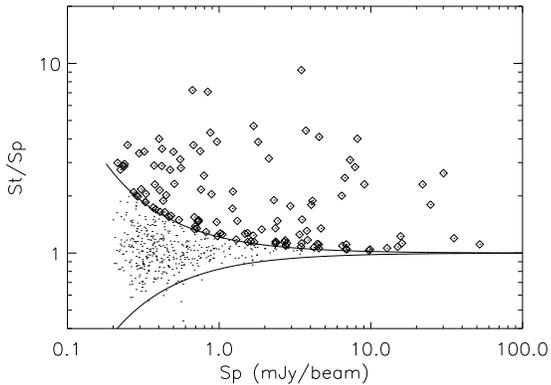}
\caption[]{Ratio of the total flux S$_{\rm T}$ to the peak flux 
S$_{\rm P}$ as a function of S$_{\rm P}$. The solid lines
show the upper and lower envelopes of the flux ratio distribution 
containing the sources considered unresolved (see text). Open symbols 
show the sources considered extended.} 
\label{ratio_fit_allflux}
\end{figure}
In Fig.~\ref{ratio_fit_allflux} we plot the ratio  between
the total S$_{\rm T}$ and the peak S$_{\rm P}$ flux density as a function of 
the peak flux for all the radio sources in the catalogue. 
To select the resolved sources, we have determined the lower envelope of 
the flux ratio distribution of Fig.~\ref{ratio_fit_allflux} and, 
assuming that values of
S$_{\rm T}$/S$_{\rm P}$ smaller than 1 are purely due to statistical errors, 
we have mirrored it above the  S$_{\rm T}$/ S$_{\rm P}$=1 value (upper 
envelope in Fig.~\ref{ratio_fit_allflux}). We have considered extended the 
116 sources lying above the upper envelope that 
can be characterized by the equation:

\begin{equation}
\frac{S_{\rm T}} {S_{\rm P}} = 0.85^{-(1.2/{\rm S_P})}~.
\end{equation}

\section{The source counts at 610 MHz}
\label{counts}
The catalogue is listed in Table~1, available in electronic form at the CDS
(http://cdsweb.u-strasbg.fr/cgi-bin/qcat?J/A+A/) or through the VVDS radio 
web page (http://virmos.bo.astro.it/radio/catalogue.html).
For each entry we list
the source name, position in RA and DEC with errors, peak brightness and
total flux density with errors,
major and minor axis, and position angle (measured from N through E).
The errors were calculated as in Paper I following the relations given
by Condon (1997)\nocite{Cond97}.

\onltab{1}{
\begin{sidewaystable*}
\begin{minipage}[t][180mm]{\textwidth}
\caption{610 MHz $5\sigma$ Catalogue: Sample page}
\label{sample_cat}
\centering
\begin{tabular}{lccccrcrcrrr}
\hline\hline
Name &  RA   &   DEC  & $\sigma_{\rm RA}$&$\sigma_{\rm DEC}$&S$_P$&
$\sigma_{\rm S_P}$& S$_T$& $\sigma_{\rm S_T}$&$\theta_M$ &$\theta_m$& PA \\
     &(J2000)&J(2000) &  ``    &   ``   & mJy/beam& mJy/beam& mJy & mJy &
     ``  & ``  & deg \\
\hline

 VIRMOS0.6GHz\_J022400-041902 & 02 24 00.24& -04 19 02.5& 0.64 & 0.55&
 0.475 &  0.077 &  0.475 &  0.077 &       &      &     \\
 VIRMOS0.6GHz\_J022400-044950 & 02 24 00.89& -04 49 50.8& 0.53 & 0.50&
 0.806 &  0.090 &  0.806 &  0.090 &       &      &     \\
 VIRMOS0.6GHz\_J022402-040731 & 02 24 02.23& -04 07 31.9& 0.62 & 0.62&
 0.759 &  0.080 &  1.641 &  0.173 & 12  &  7 &  45 \\
 VIRMOS0.6GHz\_J022402-041342 & 02 24 02.73& -04 13 42.2& 1.05 & 0.58&
 0.507 &  0.079 &  1.175 &  0.183 & 12  &  7 & 101 \\
 VIRMOS0.6GHz\_J022402-042059 & 02 24 02.08& -04 20 59.6& 0.79 & 0.75&
 0.348 &  0.069 &  0.348 &  0.069 &       &      &     \\
 VIRMOS0.6GHz\_J022402-042155 & 02 24 02.56& -04 21 55.4& 0.50 & 0.44& 
 0.823 &  0.070 &  0.823 &  0.070 &       &      &     \\
 VIRMOS0.6GHz\_J022402-044136 & 02 24 02.68& -04 41 36.7& 0.64 & 0.91&
 0.841 &  0.072 &  5.961 &  0.511 & 22  & 14 &  7  \\
 VIRMOS0.6GHz\_J022403-040931 & 02 24 03.14& -04 09 31.7& 0.65 & 0.62&
 0.750 &  0.073 &  2.584 &  0.252 & 12 & 11 & 119 \\
 VIRMOS0.6GHz\_J022403-041339 & 02 24 03.80& -04 13 39.5& 0.41 & 0.41&
 2.352 &  0.076 &  2.640 &  0.086 &  7  &  6 &  73 \\
 VIRMOS0.6GHz\_J022403-043305 & 02 24 03.80& -04 33 05.2& 0.41 & 0.41&
 4.124 &  0.084 &  7.790 &  0.158 &  9  &  8 &  92 \\
 VIRMOS0.6GHz\_J022404-044811 & 02 24 04.13& -04 48 11.2& 0.64 & 0.66&
 0.435 &  0.080 &  0.435 &  0.080 &       &      &     \\
 VIRMOS0.6GHz\_J022405-040950 & 02 24 05.31& -04 09 50.9& 0.49 & 0.44&
 0.715 &  0.069 &  0.715 &  0.069 &       &      &     \\
 VIRMOS0.6GHz\_J022405-042237 & 02 24 05.83& -04 22 37.4& 0.58 & 0.55& 
 0.341 &  0.068 &  0.341 &  0.068 &       &      &     \\
 VIRMOS0.6GHz\_J022406-045544 & 02 24 06.09& -04 55 44.6& 0.72 & 0.93&
 0.442 &  0.087 &  0.442 &  0.087 &       &      &     \\
 VIRMOS0.6GHz\_J022408-041134 & 02 24 08.87& -04 11 34.1& 0.43 & 0.43&
 1.160 &  0.068 &  1.160 &  0.068 &       &      &     \\
 VIRMOS0.6GHz\_J022408-042627 & 02 24 08.36& -04 26 27.8& 0.85 & 0.87&
 0.379 &  0.068 &  0.379 &  0.068 &       &      &     \\
 VIRMOS0.6GHz\_J022409-041702 & 02 24 09.77& -04 17 02.6& 0.45 & 0.44&
 0.918 &  0.067 &  0.918 &  0.067 &       &      &     \\
 VIRMOS0.6GHz\_J022409-044254 & 02 24 09.02& -04 42 54.7& 0.49 & 0.53&
 0.566 &  0.079 &  0.566 &  0.079 &       &      &     \\
 VIRMOS0.6GHz\_J022409-044722 & 02 24 09.94& -04 47 22.1& 0.59 & 0.49&
 0.579 &  0.077 &  0.579 &  0.077 &       &      &     \\
 VIRMOS0.6GHz\_J022410-042147 & 02 24 10.03& -04 21 47.7& 0.46 & 0.44&
 0.788 &  0.064 &  0.788 &  0.064 &       &      &     \\
 VIRMOS0.6GHz\_J022410-044607A& 02 24 09.58& -04 46 06.4& 0.40 & 0.40&
 8.943 &  0.089 & 13.472 &  0.134 &  8  &  7 & 102 \\
 VIRMOS0.6GHz\_J022410-044607B& 02 24 10.36& -04 46 08.2& 0.40 & 0.40&
 24.735&  0.086 & 30.740 &  0.107 &  7  &  6 &  96 \\
 VIRMOS0.6GHz\_J022410-044607T& 02 24 10.13& -04 46 07.5& 0.40 & 0.40&
 24.735&  0.058 & 44.500 &  0.104 & 30  & 15 &      \\
 VIRMOS0.6GHz\_J022410-045033 & 02 24 10.22& -04 50 33.5& 0.40 & 0.40&
 6.859 &  0.082 &  7.207 &  0.086 &  6  &  6 &  0  \\
 VIRMOS0.6GHz\_J022411-042604 & 02 24 11.55& -04 26 04.8& 0.95 & 1.17&
 0.326 &  0.062 &  0.606 &  0.116 & 13  &  6 &  37  \\
 VIRMOS0.6GHz\_J022412-041042 & 02 24 12.27& -04 10 42.0& 0.43 & 0.42&
 1.065 &  0.064 &  1.065 &  0.064 &       &      &     \\
 VIRMOS0.6GHz\_J022412-044043 & 02 24 12.34& -04 40 43.5& 0.47 & 0.52&
 0.462 &  0.071 &  0.462 &  0.071 &       &      &     \\
 VIRMOS0.6GHz\_J022413-042227 & 02 24 13.63& -04 22 27.6& 0.60 & 0.53&
 0.551 &  0.063 &  0.551 &  0.063 &       &      &     \\
 VIRMOS0.6GHz\_J022413-042732 & 02 24 13.63& -04 27 32.1& 0.49 & 0.47&
 0.798 &  0.063 &  0.798 &  0.063 &       &      &     \\
 VIRMOS0.6GHz\_J022413-044642 & 02 24 13.18& -04 46 42.2& 0.42 & 0.42&
 1.484 &  0.076 &  1.484 &  0.076 &       &      &     \\
 VIRMOS0.6GHz\_J022414-041205 & 02 24 14.24& -04 12 05.7& 0.64 & 0.53&
 0.386 &  0.065 &  0.386 &  0.065 &       &      &     \\
 VIRMOS0.6GHz\_J022414-041237 & 02 24 14.16& -04 12 37.0& 0.64 & 0.59&
 0.339 &  0.066 &  0.339 &  0.066 &       &      &     \\
 VIRMOS0.6GHz\_J022414-042922 & 02 24 14.57& -04 29 22.9& 0.94 & 0.52&
 0.338 &  0.067 &  0.338 &  0.067 &       &      &     \\
 VIRMOS0.6GHz\_J022415-043143 & 02 24 15.98& -04 31 43.8& 0.41 & 0.41&
 1.640 &  0.064 &  1.640 &  0.064 &       &      &     \\
 VIRMOS0.6GHz\_J022416-042056 & 02 24 16.07& -04 20 56.8& 0.50 & 0.46&
 0.591 &  0.060 &  0.591 &  0.060 &       &      &     \\
 VIRMOS0.6GHz\_J022419-042028 & 02 24 19.04& -04 20 28.2& 0.62 & 0.55&
 0.319 &  0.059 &  0.319 &  0.059 &       &      &     \\
\hline
\end{tabular}
\vfill
\end{minipage}
\end{sidewaystable*}
}

\begin{figure}
\includegraphics[width=8cm]{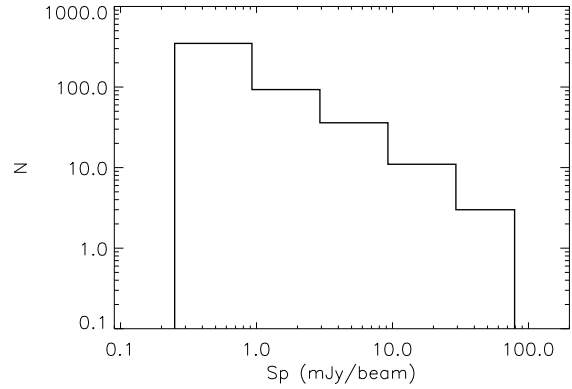}
\caption[]{Peak flux density distribution for all the 514 radio sources at
610 MHz}
\label{Speak_distribution}
\end{figure}

For the unresolved sources the total flux density is set equal to the peak
brightness found by MAXFIT and the angular size is undetermined.
For each of the 17 sources fitted with multiple components in the
catalogue we list an entry for each of the components, identified with a 
trailing 
letter (A, B, C, {\ldots}) in the source name, and an entry for the whole
source, identified with a trailing T in the source name. 
In these cases the total flux was calculated 
using the task TVSTAT, which allows for the integration of the map values over
irregular areas, and the sizes are the largest angular sizes. 
The peak flux density distribution
of the 514 radio sources is shown in Fig.~\ref{Speak_distribution}.
The catalogue contains 381 radio sources with a flux density less than 1 mJy 
at 610 MHz (corresponding to $74\%$ of the whole sample) and, to our 
knowledge, it is the deepest survey at low frequencies 
($\nu < 1.4$ GHz) obtained so far.

To minimize incompleteness effects in the lowest flux density bin we derived 
the radio counts using the radio sources with flux
density greater than 0.3 mJy. The source counts are summarised in
Table~\ref{Tab_counts}.
\begin{table*}
\centering
\caption{The 610 MHz radio source counts}
\label{Tab_counts}
     
\begin{tabular}{cccccccc}
     
\hline

$S$  & $<S>$ & $N_S$ & $N_S^{\rm corr}$ & $dN/dS$             & $nS^{2.5}$          &  $N(>S)$ &C$_{\rm res}$\\
(mJy) & (mJy) &       &                 &sr$^{-1}$ Jy$^{-1}$ & sr$^{-1}$ Jy$^{1.5}$ &deg$^{-2}$&           \\
\hline
~~0.30 -- ~~0.45 &   0.37 & 145 & 195.4 & 4.28$\times10^9$ & ~11.06$\pm$~~0.93   &488 & 1.5 \\
~~0.45 -- ~~0.68 &   0.55 &  93 &  93.4 & 1.36$\times10^9$ & ~~9.72$\pm$~~1.01   &292 & 1.3\\
~~0.68 -- ~~1.01 &   0.83 &  64 &  64.0 & 6.23$\times10^8$ & ~12.23$\pm$~~1.53   &199 & 1.2\\
~~1.01 -- ~~1.52 &   1.24 &  33 &  33.0 & 2.14$\times10^8$ & ~11.59$\pm$~~2.02   &135 & 1.1\\
~~1.52 -- ~~2.28 &   1.86 &  32 &  32.0 & 1.38$\times10^8$ & ~20.64$\pm$~~3.65   &102 & 1.0\\
~~2.28 -- ~~3.42 &   2.79 &  19 &  19.0 & 5.48$\times10^7$ & ~22.52$\pm$~~5.17   &~70 & 1.0\\
~~3.42 -- ~~5.13 &   4.19 &  16 &  16.0 & 3.07$\times10^7$ & ~34.84$\pm$~~8.71   &~51 & 1.0\\
~~5.13 -- ~~7.69 &   6.28 &  11 &  11.0 & 1.41$\times10^7$ & ~44.00$\pm$~13.27   &~35 & 1.0\\
~~7.69 -- ~11.53 &   9.42 &   4 &   4.0 & 3.42$\times10^6$ & ~29.39$\pm$~14.70   &~24 & 1.0\\
~11.53 -- ~17.30 &  14.13 &   6 &   6.0 & 3.42$\times10^6$ & ~80.99$\pm$~33.07   &~20 & 1.0\\
~17.30 -- ~25.95 &  21.19 &   6 &   6.0 & 2.28$\times10^6$ & 148.80$\pm$~60.75   &~14 & 1.0\\
~25.95 -- ~38.92 &  31.78 &   2 &   2.0 & 5.06$\times10^5$ & ~91.12$\pm$~64.43   &~~8 & 1.0\\
~38.92 -- ~58.39 &  47.67 &   4 &   4.0 & 6.75$\times10^5$ & 334.79$\pm$167.40   &~~6 & 1.0\\
~58.39 -- ~87.59 &  71.51 &   1 &   1.0 & 1.13$\times10^5$ & 153.76$\pm$153.76   &~~2 & 1.0\\
~87.59 -- 131.37 & 107.26 &   0 &   0.0 & 0.00             & 0.00                &~~1 & 1.0\\
131.37 -- 197.05 & 160.89 &   1 &   1.0 & 5.00$\times10^4$ & 518.95$\pm$518.95   &~~1 & 1.0\\
\hline
\end{tabular}
\end{table*}
For each flux density bin we report the average flux density, the observed
number of sources, the number of sources corrected for the visibility area,
the differential source density (in sr$^{-1}$ Jy$^{-1}$),
the normalised differential counts $nS^{2.5}$ (in sr$^{-1}$ Jy$^{1.5}$) with
estimated Poissonian errors (as $n^{1/2}S^{2.5}$), the cumulative number 
of sources, and the resolution bias correction.
The resolution bias correction accounts for the fact that extended objects
with peak flux densities below the survey limit, but total integrated flux 
densities above this limit, would not be detected by the search procedure.
Given a flux density, there will be a maximum detectable angular size beyond
which a source will no longer be detectable because its peak flux density
drops below the catalogue threshold.
To estimate the correction factors to be applied we modeled a population of
radio sources extracted from source counts with the integral size distribution
derived in Paper I and described by a broken power law consistent with that
observed.

\begin{figure}
\includegraphics[width=9cm]{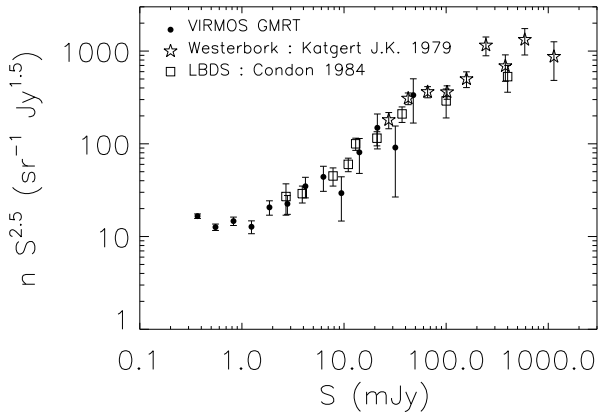}
\caption[]{The normalized differential source counts at 610 MHz. Our results
(dots) are shown together with those from 2 other surveys
(\cite{Katg79}; \cite{Cond84}).}
\label{Source_counts}
\end{figure}

The normalized differential counts are shown in
Fig.~\ref{Source_counts} together with other measurements obtained at the
same frequency from different surveys. Even if not comparable to the deepness
of radio surveys at higher frequencies, the source counts obtained for the
F02 field at 610 MHz are about one order of magnitude deeper than any
previous survey below 1 GHz.
%The counts in Table~\ref{Tab_counts} and Figure~\ref{Source_counts}
%have only been corrected for the visibility area but not for any other 
%possible bias, such as the resolution bias or the effect of noise 
%on the source extraction and flux density determination. A detailed
%description of a proper way to take these effects into account is given 
%in Paper I for the 1.4 GHz radio source counts of the F02 field
%and a similar analysis was considered beyond the purpose 
%of the present work. Furthermore, it is worth reminding that since the 
%resolution of the 1.4 GHz and 610 MHz observations and the catalogue
%extraction procedures for the two surveys were the same we should expect 
%similar incompletness factors as a function of the signal to noise ratio
%between the two surveys (roughly of the order of $15-20\%$ in the first two
%flux density bins and no correction at higher fluxes).
Figure~\ref{Source_counts} clearly shows a flattening of the radio counts 
around 1 mJy at 610 MHz. 
%Such effect would be increased once all the 
%incompletness factors are taken into account.

\subsection{Comparison with the 1.4 GHz radio source catalogue}

Both the 610 MHz and 1.4 GHz catalogues have been derived with the same 
procedure from images with the same resolution but different sensitivity
(the r.m.s. in the 1.4 GHz mosaic is $\simeq$ 17 $\mu$Jy). 
The 610 MHz catalogue contains 514 radio sources with flux density greater
than 200 $\mu$Jy compared with the 1054 radio sources with flux density
greater than 71 $\mu$Jy in the 1.4 GHz catalogue.
We used all the unresolved sources with flux greater than 0.5 mJy at 610 MHz
to derive the systematic offset in right ascension and declination between
the two catalogues. Assuming the 1.4 GHz VLA catalogue as a reference, we
find a mean offset of $\Delta$RA$=-0.14\pm 0.37$ and $\Delta$DEC$=0.87\pm 0.40$
arcsec. 
These offsets are a fraction of the pixel size and they are expected since the 
the data sets were self-calibrated.
Taking into account these systematic offsets we cross-correlated
the two $5\sigma$ catalogues using a search radius of 3 arcsec (2 pixels). 
We found that 
448 sources (87\%) in the 610 MHz catalogue have a matched counterpart at 1.4 
GHz with only 66 sources detected at 610 MHz without a counterpart above
$5\sigma$ at 1.4 GHz.
Given the different sensitivity of the two surveys and the frequency range
considered, these 66 sources will have spectral index steeper than $\sim -1$.
We will return to these sources later in Sect. 6.

\section{The spectral index properties of sub-mJy radio sources}

\begin{figure}
\includegraphics[width=8cm]{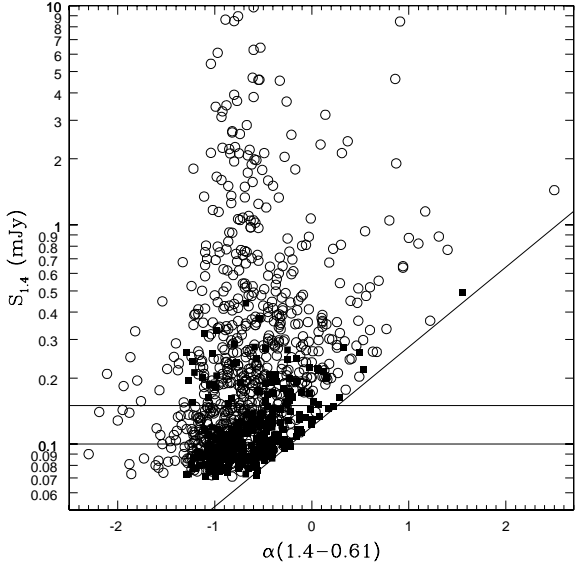}
\caption[]{Flux density at 1.4 GHz as a function of the spectral index
between 1.4 GHz and 610 MHz. Filled squares represent lower limits in the
spectral index. The solid lines, drawn at 0.10 and 0.15 mJy, are used to
illustrate the bias against flat spectrum radio sources at lower flux levels.
The slanted solid line shows the maximum value of the spectral index above
which no 610 MHz detection is possible due to the flux limit of the 610
MHz survey.}
\label{flux_vs_spix}
\end{figure}
\subsection{The spectral index catalogue}

Both the 610 MHz and 1.4 GHz surveys have the same angular resolution and the
catalogues have been assembled using identical procedures. These features
are important to derive reliable spectral indeces.
Moreover, the large number of sub-mJy sources found in these surveys
allows us to study the spectral index properties of this class of sources with
robust statistics.

The present analysis assumes that the radio sources have not varied
significantly in the 4 year interval between the 1.4 GHz and 610 MHz
observations. For the general properties of the sample this is a reasonable 
assumption since a significant variability
occurs in only about 5\% of the mJy and sub-mJy radio source population
(\cite{OW85}; \cite{Wind85}).
To study the spectral index distribution of sub-mJy radio sources
we started from the 1.4 GHz $5\sigma$ catalogue searching for counterparts at 
610 MHz within a radius of 3 arcseconds 
from the position of the component at 1.4 GHz 
and to a limit of $3\sigma$ at 610 MHz.
Of the 1054 radio sources in the 1.4 GHz catalogue, 448 are coincident
with radio sources above $5\sigma$ at 610 MHz, 293 correspond to sources
in the range 3-5$\sigma$ at 610 MHz, and 313 have no counterpart
above $3\sigma$ at the lower frequency. 
The spectral index catalogue is available as an ascii table at the following
address: 
http//virmos.bo.astro.it/radio/catalogue.html .
Spectral indeces (defined as $S\propto \nu^\alpha$) for the 741 matched 
radio sources (with counterparts above $3\sigma$ at 610 MHz) are calculated 
using the total radio flux density at each frequency. Lower limits are
derived for the 313 sources without a counterpart at 610 MHz using the 
peak brightness of the 1.4 GHz component and 3 times the local rms in 
the 610 MHz image.

\subsection{Spectral index versus radio flux density}
In Fig.~\ref{flux_vs_spix} the 1.4 GHz flux density versus the radio
spectral index is shown for the 1054 
sources (detection and limits).
Filled squares are used to indicate the lower limits in the
spectral index.  
Figure~\ref{flux_vs_spix} clearly shows the well known bias in spectral index
due to the different sensitivities of the observations at the two frequencies. 
In our case, there
is a bias against flat spectrum sources at lower flux densities because of the
higher flux limit of the 610 MHz observations. For illustrative
purposes, we drew two lines corresponding to $S$(1.4 GHz)$=0.10$ mJy and 
$S$(1.4 GHz)$=0.15$ mJy. Below 0.1 mJy only the sources with the 
steepest spectral index can be detected at 610 MHz, and the lower limits 
dominate 
(108 detections and 134 lower limits for sources with $S$(1.4 GHz)$<0.10$ mJy).
Therefore to avoid such a strong bias, 
in the following statistical analysis we have considered only the 812 
spectral indeces (633 detections and 179 lower limits) derived from sources 
with $S$(1.4 GHz)$\ge 0.10$ mJy.

\begin{table}
 \centering
 \caption{The radio spectral index versus 1.4 GHz flux}
 \label{tab_spix_flux}
     \begin{tabular}{lccc}
     \hline
1.4 GHz Flux    & \# of        & \# of      & Median \\
 interval (mJy) &  detections  & limits     &  $\alpha_{\rm r\_med}$ \\
\hline
$0.10\le S < 0.15$ & 171     & 110        & $-0.61\pm 0.04$ \\
$0.15\le S < 0.50$ & 304     & ~69        & $-0.46\pm 0.03$ \\
$S\ge 0.50$         & 158     & ~~0        & $-0.67\pm 0.05$ \\
\hline
\end{tabular}
\end{table}

In our analysis we included both the measured values and the lower
limits to $\alpha_{\rm r}$.
The sources were divided into three flux density bins: $0.10\le S$(1.4
GHz)$< 0.15$ mJy, $0.15\le S$(1.4 GHz)$< 0.50$ mJy, and $S$(1.4 GHz)$\ge
0.50$ mJy, and for each bin we calculated the median value of the
spectral index taking into account the lower limits. The statistical
difference of the median in the three flux density intervals was tested using 
the software package ASURV, which
implements the methods described by \cite{FN85} and \cite{IFN86} to properly 
handle censored data.
The radio spectral index of the first and third bin
are consistent with being drawn from the same distribution, while the
sources with $0.15\le S$(1.4 GHz)$< 0.50$ mJy have 99\% of probability
of being drawn from a different population with flatter spectral index.
This result, based on a much larger number of sources than previous works
is statistically significant and deserves some comments.

The well known flattening of the normalized source counts appears at 
$S$(1.4 GHz)$\la 0.5$ mJy, so we can expect that the highest flux bin is 
dominated by classical radio loud AGNs with spectral index $\sim -0.7$, 
consistent with that found.
The situation becomes more interesting at lower flux densities where,
as we already know from the radio counts, there is a rapid increase in the
number of radio sources that is generally ascribed to
starburst galaxies, low luminosity AGNs, and radio quiet QSOs.
Indeed, we detect a significantly flatter spectral index for sources with
$0.15\le S$(1.4 GHz)$< 0.50$ mJy, meaning that the new population of radio
sources emerging below 0.5 mJy has a relevant contribution from flat spectrum
compact radio cores that could be hosted in low luminosity AGNs or in radio 
quiet QSOs. On the contrary, the starburst population is expected to have 
$\alpha \sim -0.7$ or steeper, since the contribution of thermal radio emission
at these frequencies is negligible.
It is worth noting that, because of the spectral index bias, we 
eventually miss radio sources with flat or inverted spectra. 
Therefore, we can conclude that the flatter
radio spectrum of the sources in the intermediate flux density bin compared
to those at higher flux densities is statistically significant. 
Even if it can be difficult to compare results obtained by samples selected at
different frequencies, this result is 
also consistent with that found by Ciliegi et al. (2003)\nocite{Cili03} 
in the Lockman Hole Survey, where radio sources with flux density at 4.9 GHz
in the range $0.1\le S<0.2$ mJy have $\alpha_{\rm r\_med}=-0.37\pm 0.10$
mJy compared to $\alpha_{\rm r\_med}=-0.81\pm 0.14$ mJy for the sources at
higher flux density.

The interpretation of the spectral index distribution in the fainter flux 
density bin  ($0.10\le S$(1.4 GHz)$< 0.15$ mJy) is more uncertain due to 
the increasing weight of the bias against flat spectrum 
sources. The value of $\alpha_{\rm r\_med}=-0.61\pm 0.04$ mJy
suggests that the contribution from flatter spectrum radio cores
is decreasing and steeper spectra sources are taking over again. Indeed,
evolutionary models of the sub-mJy population predict that starburst
galaxies become dominant at $S\la 0.2$ mJy at 1.4 GHz (e.g., \cite{Hopk98}; 
\cite{SMG04}), and the spectral index properties of sources below 0.15 mJy 
might just be the signature of this effect. Unfortunately, the large number 
of lower limits in the first flux density bin ($\simeq 40\%$) and the 
selection effect against sources with spectral index flatter than $\simeq 0$ 
that is still present in the first flux density bin make it difficult to draw 
any strong conclusion.

In summary,
the spectral index distribution shows statistically significant differences 
above and below 0.5 mJy at 1.4 GHz, confirming that a change in the population
of radio sources occurs below 0.5 mJy. Radio sources between 0.5 mJy and
0.15 mJy have on 
average a flatter spectral index, and we interpret this result as being
due to the contribution of flat spectrum cores in low luminosity AGNs and 
radio quiet QSOs. At the faintest flux density level of our survey, 
the spectral index steepens again, as expected if starburst galaxies become
the dominant population, but possible incompleteness effects
and the large number of limits makes the statistical significance of this
result less secure.

\subsection{Spectral index versus optical type}

Optical identifications of the 1.4 GHz catalogue were presented in Paper II.
On the basis of a maximum likelihood technique, 718 of the 1054
radio sources were identified with optical counterparts and classified in
different optical types based on a color-color plot.
We adopted a revised classification of the optical type of the identified
radio sources based on the best-fit SED template
used to estimate the photometric redshift (details on the template sets
used to fit the photometric data can be found in \cite{Ilbe06}). 
Due to the uncertainties of the
process, we decided to use only 3 main morphological categories 
(early-type, late-type, and starburst) and considered only objects whose 
photometric redshifts
are in range $0.2\le z\le 1.5$, where there is an excellent agreement between
photometric and spectroscopic redshifts in the VIMOS-VLT deep survey
(\cite{Ilbe06}). The results of the spectral index distribution for
different optical types are shown in Table~\ref{tab_spix_type}.

\begin{table}
 \centering
 \caption{The radio spectral index versus optical type}
 \label{tab_spix_type}
     \begin{tabular}{lcccc}
     \hline
Optical        &  \# of      & \% of flat & \# of  & Median \\
classif.       &  detections  &    spectrum& limits &  $\alpha_{\rm r\_med}$ \\
\hline
Early          & 225         & 34\%      &  65     & $-0.55\pm 0.04$ \\
Late           & ~80         & 21\%      &  21     & $-0.70\pm 0.04$ \\
Starburst      & ~37         & 11\%      &  ~8     & $-0.69\pm 0.10$ \\
Unidentified   & 263         & 34\%      &  64     & $-0.59\pm 0.04$ \\
\hline
\end{tabular}
\end{table}

A statistical analysis of the spectral index distribution for each optical
classification was performed with ASURV.
The median spectral index of the early-type galaxies is significantly flatter
than that of late-type objects. Based on all the test of the ASURV analysis, 
there is a 98\% probability that the two distributions are different.
Early-type and starburst galaxies have also different $\alpha_{\rm r\_med}$
with a slightly less significance because of the larger uncertainty. 
The unidentified objects have a spectral index distribution indistinguishable
from early-type galaxies and with a probability of 95\% of being different from
that of late-type objects. Finally, the fraction of flat spectrum radio sources
($\alpha>-0.5$) is the same in early-type and unidentified objects, while it
is significantly lower in late-type and starburst ones. 
These results confirm that 
late-type and starburst galaxies have steeper radio spectra than radio sources 
identified with early-type objects. The radio spectral properties of the 
unidentified radio sources suggest that
a large fraction of optically unidentified objects is associated with
weak (distant or obscured) early-type galaxies.

\section{Steep spectrum radio sources and high-redshift radio galaxies}

While it is observationally well known that high-redshift radio galaxies
(HzRGs) have the steepest spectral indices (usually measured between
a few hundred MHz and 1.4 GHz), the reasons why this happens are not entirely
understood. Several explanations have been invoked in the literature to
interpret this so-called $z$-$\alpha$ correlation. Most of them assume that the
spectral energy distribution of high and low redshift radio galaxies are 
intrinsically identical and that $k$-correction and enhanced inverse Compton  
steepen the spectra of HzRGs (De Breuck et al. 2000, 
2002\nocite{DeBr00}\nocite{DeBr02}; \cite{Peda03}; \cite{Cohe04}; 
\cite{Jarv04}).
Recently, the possibility that HzRGs could have 
intrinsically steeper spectral indices as they might expand in denser 
environments has 
been suggested to explain the radio SED in the SUMSS-NVSS sample
of ultra-steep spectrum sources (\cite{Klam06}).

We used our dual frequency observations of the F02 field to derive a
sample of ultra-steep spectrum (USS) radio sources 
selecting all the radio sources with $\alpha \le -1.3$. To minimize the
inclusion of spurious sources (e.g., noise spikes), 
we considered only those sources with a
$\ge 5\sigma$ detection at one frequency and at least a $\ge 3\sigma$
detection at the other frequency. Finally, to avoid the contribution 
of steep spectrum nearby sources we excluded all the sources found resolved 
in the 1.4 GHz catalogue.

The sample contains
39 objects from the 1.4 GHz based spectral index catalogue (i.e. with a
$\ge 5\sigma$ detection at 1.4 GHz, and $\ge 3\sigma$ counterpart at 610 MHz). 
To this sample we added the 19 sources from the complete 610 MHz sample ($\ge
5\sigma$) for which there is a counterpart between 3 and 5 $\sigma$ at 1.4 
GHz. This sample is listed in Table~\ref{tab_uss}, with the name from the
1.4 GHz catalogue, or from the 610 MHz catalogue for those objects with
1.4 GHz flux between 3 and 5 $\sigma$, total flux at 1.4 GHz, peak brightness
and total flux at 610 MHz, relative errors, and the spectral index. 

These 58 sources must be considered as candidate USS sources
as several effects can be responsible for the steep spectral index
regardless of the intrinsic shape of the radio spectrum. 
Since the 1.4 GHz and
610 MHz observations are separated by 4 years, radio flux density variability 
may contaminate the sample. 
Furthermore, all the 58 USS candidates
are weak radio sources: the brightest one is 0.33 mJy at 1.4 GHz. This means 
that 
errors in the total flux density determination can also be relatively large,
yielding to a less secure spectral index value.
Finally, larger fitted sizes for some components at 610 MHz compared to the
sizes derived at 1.4 GHz can produce
very steep spectral indices. This last effect is certainly affecting some
of the sources with the most extreme spectral index ($\alpha < -1.9$) which 
are fitted with extended Gaussian components at
610 MHz. Anyway, it must be noted that for these sources even considering the
peak flux density at 610 MHz instead of the total flux density yields to
sources with steep spectral index ($\alpha \la -1$).

Considering all these effects and possible contaminations, the sample of
58 candidate USS sources is still a good starting point to 
select real ultra-steep spectrum objects and simultaneous multifrequency
observations are needed to provide a reliable classification. 
In particular, it is interesting to
note how these candidates are about one to two orders of magnitude fainter
than USS sources selected from previous studies (e.g., 
De Breuck et al. 2000, 2004\nocite{DeBr00}\nocite{DeBr04}) 
so they could be associated with the most distant USS sources or could be the
low radio luminosity counterparts of the HzRGs presently known. 

\begin{figure}
\includegraphics[width=8cm]{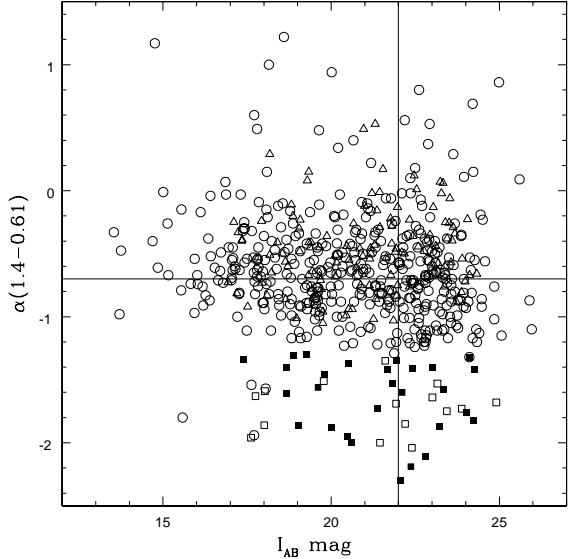}
\caption[]{Two point radio spectral index as a function of the optical
I$_{\rm AB}$ magnitude. Dots are used for sources in the spectral index 
catalogue and not USS, triangles are for lower limits in the  
spectral index catalogue and not USS, filled squares
are USS candidates in the catalogue and empty squares are USS candidates
not in the catalogue (with 1.4 GHz counterpart between 3 and 5 $\sigma$).}
\label{spix_iab}
\end{figure}

Among the 58 USS candidates 45 are optically identified and we tried to 
confirm the nature of possible HzRGs using the available optical information.
In Fig.~\ref{spix_iab} we plotted the I$_{\rm AB}$ magnitude and spectral index
for all the sources using the following notation: empty circles are sources
in the spectral index catalogue and not USS candidates, triangles are 
sources in the spectral index catalogue and not USS candidates 
with a lower limit in the spectral index, filled squares are 
USS candidates that also belong to the spectral index catalogue, and finally
empty squares are used to identify USS candidates that do not belong to the
spectral index catalogue (sources with 1.4 counterparts between 3
and 5 $\sigma$).
The small number of USS candidates makes a statistical comparison rather
uncertain, but the median of the I$_{\rm AB}$ magnitude for the USS candidates
is 21.8 compared to 20.9 for the rest of the sample.
What can be appreciated from Fig.~\ref{spix_iab} is a general steepening of the
spectral index going to fainter magnitudes. 
We have divided all the identified radio sources into two bins of 
optical I$_{\rm AB}$ magnitude, smaller or greater than I$_{\rm AB}=22$, 
and the results are shown in  Table~\ref{tab_spix_iab}.
The two distributions are different with a confidence level of 99\%.

\begin{table}
 \centering
 \caption{The radio spectral index versus optical magnitude}
 \label{tab_spix_iab}
     \begin{tabular}{lccc}
     \hline
 I$_{\rm AB}$     & \# of   & \# of  & Median \\
    mag           &  detections  & limits     &  $\alpha_{\rm r\_med}$ \\
\hline
I$_{\rm AB}<22$         & 351         &  80        & $-0.57\pm 0.03$ \\
I$_{\rm AB}\ge 22$      & 193         &  33        & $-0.68\pm 0.04$ \\
\hline
\end{tabular}
\end{table}
A limited area (625 arcmin$^2$) of the F02 field was observed in the  
near-infrared K-band optical (\cite{Iovi05}; \cite{Temp07}).
Twelve of the 58 USS candidates fall in this area; 11 out of 12 have an 
optical/near-infrared
identification and 8 out of 11 (73\%) have K$_{\rm Vega}>$ 20.
For comparison, 120 sources from the spectral index catalogue and not USS
are identified in the same area and only 65 out of 120 (54\%) have 
K$_{\rm Vega}> $20. Clearly the statistics are still very limited, but this
is another indication of a larger fraction of possible HzRGs in the USS 
sources compared to the rest of the sample.
An estimate of photometric redshifts through the best-fit
to the optical/near-infrared spectral energy distribution, following the
method described in Ilbert et al. (2006)\nocite{Ilbe06}, indicates that 
these 11 sources are in the redshift range $\sim$ 0.2 -- 3.5, with only 3 of them having
photometric redshift $<$ 1. The observed spectral energy distributions
of these sources are consistent with a variety of spectral types,
ranging from elliptical to irregular/starburst. A closer look at their
optical and K-band images reveals a high incidence of objects with
peculiarities, while the position of the sources in the BzK diagram
introduced by \cite{Dadd04} are consistent with 3 of these sources 
being z $>$ 1.4 star-forming galaxies and 1 a z $>$ 1.4 passive galaxy, 
which also has R$-$K and I$-$K colors typical of extremely red objects.

\begin{table*}
\label{tab_uss}
\caption{Candidate USS sources}
\centering
\begin{tabular}{lccccccc}
\hline
\hline
     &  \multicolumn{2}{c}{1.4 GHz}& \multicolumn{4}{c}{0.61 GHz} & \\
Name & S$_T$ & $\sigma_{\rm S_T}$ & S$_p$ &
$\sigma_{\rm S_p}$ & S$_T$ & $\sigma_{\rm S_T}$ & 
$\alpha$ \\
     & mJy           &     mJy                    & mJy/b.      & 
 mJy/b.                   &    mJy         &           mJy               &
         \\
\hline
 VIRMOS1.4GHz\_J022402-041343 &  0.249   &  0.017   & 0.507    & 0.079    & 1.175    & 0.183   & -1.87\\
 VIRMOS1.4GHz\_J022406-044556 &  0.104   &  0.018   & 0.332    & 0.086    & 0.332    & 0.086   & -1.40\\
 VIRMOS1.4GHz\_J022411-042604 &  0.090   &  0.018   & 0.326    & 0.062    & 0.606    & 0.116   & -2.30\\
 VIRMOS1.4GHz\_J022413-044642 &  0.327   &  0.020   & 1.484    & 0.076    & 1.484    & 0.076   & -1.82\\
 VIRMOS1.4GHz\_J022414-041204 &  0.081   &  0.015   & 0.386    & 0.065    & 0.386    & 0.065   & -1.88\\
 VIRMOS1.4GHz\_J022422-042616 &  0.157   &  0.016   & 0.329    & 0.066    & 0.675    & 0.135   & -1.76\\
 VIRMOS1.4GHz\_J022423-040229 &  0.128   &  0.016   & 0.408    & 0.073    & 0.673    & 0.121   & -2.00\\
 VIRMOS1.4GHz\_J022428-041511 &  0.073   &  0.015   & 0.342    & 0.058    & 0.342    & 0.058   & -1.86\\
 VIRMOS1.4GHz\_J022432-041342 &  0.250   &  0.016   & 0.543    & 0.061    & 0.813    & 0.092   & -1.42\\
 VIRMOS1.4GHz\_J022436-041031 &  0.139   &  0.016   & 0.306    & 0.061    & 0.665    & 0.133   & -1.88\\
 VIRMOS1.4GHz\_J022436-041709 &  0.097   &  0.016   & 0.288    & 0.053    & 0.288    & 0.053   & -1.31\\
 VIRMOS1.4GHz\_J022444-044335 &  0.096   &  0.017   & 0.320    & 0.053    & 0.320    & 0.053   & -1.45\\
 VIRMOS1.4GHz\_J022447-045851 &  0.133   &  0.017   & 0.418    & 0.073    & 0.418    & 0.073   & -1.38\\
 VIRMOS1.4GHz\_J022451-045700 &  0.143   &  0.018   & 0.439    & 0.082    & 0.722    & 0.135   & -1.95\\
 VIRMOS1.4GHz\_J022452-040912 &  0.084   &  0.015   & 0.248    & 0.049    & 0.248    & 0.049   & -1.30\\
 VIRMOS1.4GHz\_J022517-044602 &  0.084   &  0.016   & 0.270    & 0.051    & 0.270    & 0.051   & -1.41\\
 VIRMOS1.4GHz\_J022550-042141 &  0.135   &  0.016   & 0.414    & 0.044    & 0.414    & 0.044   & -1.35\\
 VIRMOS1.4GHz\_J022550-044505 &  0.122   &  0.019   & 0.372    & 0.049    & 0.372    & 0.049   & -1.34\\
 VIRMOS1.4GHz\_J022611-044005 &  0.078   &  0.015   & 0.289    & 0.045    & 0.289    & 0.045   & -1.58\\
 VIRMOS1.4GHz\_J022621-040834 &  0.140   &  0.017   & 0.377    & 0.059    & 0.867    & 0.136   & -2.19\\
 VIRMOS1.4GHz\_J022624-044203 &  0.209   &  0.017   & 0.418    & 0.049    & 1.204    & 0.140   & -2.11\\
 VIRMOS1.4GHz\_J022638-040845 &  0.090   &  0.017   & 0.270    & 0.058    & 0.270    & 0.058   & -1.32\\
 VIRMOS1.4GHz\_J022642-044205 &   0.220  &  0.017   & 0.758    & 0.046    & 0.758    & 0.046   & -1.49\\
 VIRMOS1.4GHz\_J022658-041814 &   0.217  &  0.016   & 0.683    & 0.050    & 0.683    & 0.050   & -1.38\\
 VIRMOS1.4GHz\_J022700-040501 &   0.233  &  0.014   & 0.467    & 0.066    & 0.725    & 0.103   & -1.37\\
 VIRMOS1.4GHz\_J022709-045117 &   0.080  &  0.015   & 0.305    & 0.058    & 0.305    & 0.058   & -1.61\\
 VIRMOS1.4GHz\_J022713-041756 &   0.083  &  0.016   & 0.283    & 0.054    & 0.283    & 0.054   & -1.48\\
 VIRMOS1.4GHz\_J022716-045719 &   0.086  &  0.016   & 0.363    & 0.083    & 0.363    & 0.083   & -1.73\\
 VIRMOS1.4GHz\_J022724-042502 &   0.115  &  0.016   & 0.366    & 0.053    & 0.366    & 0.053   & -1.39\\
 VIRMOS1.4GHz\_J022727-045724 &   0.288  &  0.016   & 0.689    & 0.083    & 0.921    & 0.110   & -1.40\\
 VIRMOS1.4GHz\_J022735-043159 &   0.155  &  0.016   & 0.466    & 0.057    & 0.466    & 0.057   & -1.32\\
 VIRMOS1.4GHz\_J022735-044628 &   0.118  &  0.018   & 0.447    & 0.069    & 0.447    & 0.069   & -1.60\\
 VIRMOS1.4GHz\_J022739-040504 &   0.074  &  0.013   & 0.263    & 0.067    & 0.263    & 0.067   & -1.53\\
 VIRMOS1.4GHz\_J022741-041736 &   0.084  &  0.015   & 0.312    & 0.063    & 0.312    & 0.063   & -1.58\\
 VIRMOS1.4GHz\_J022743-044721 &   0.147  &  0.018   & 0.437    & 0.073    & 0.437    & 0.073   & -1.31\\
 VIRMOS1.4GHz\_J022744-040911 &   0.103  &  0.014   & 0.347    & 0.067    & 0.347    & 0.067   & -1.46\\
 VIRMOS1.4GHz\_J022745-040559 &   0.110  &  0.015   & 0.359    & 0.070    & 0.359    & 0.070   & -1.42\\
 VIRMOS1.4GHz\_J022755-044520 &   0.100  &  0.019   & 0.360    & 0.085    & 0.360    & 0.085   & -1.54\\
 VIRMOS1.4GHz\_J022758-045432 &   0.110  &  0.020   & 0.403    & 0.102    & 0.403    & 0.102   & -1.56\\
 VIRMOS0.6GHz\_J022414-042922 &   0.062  &  0.018   & 0.338    & 0.067    & 0.338    & 0.067   & -2.04\\
 VIRMOS0.6GHz\_J022421-041713 &   0.081  &  0.017   & 0.313    & 0.061    & 0.313    & 0.061   & -1.63\\
 VIRMOS0.6GHz\_J022424-044327 &   0.060  &  0.015   & 0.315    & 0.063    & 0.315    & 0.063   & -2.00\\
 VIRMOS0.6GHz\_J022503-044807 &   0.072  &  0.018   & 0.271    & 0.048    & 0.271    & 0.048   & -1.60\\
 VIRMOS0.6GHz\_J022512-044720 &   0.068  &  0.018   & 0.292    & 0.050    & 0.292    & 0.050   & -1.75\\
 VIRMOS0.6GHz\_J022525-043235 &   0.050  &  0.015   & 0.199    & 0.039    & 0.199    & 0.039   & -1.66\\
 VIRMOS0.6GHz\_J022538-044658 &   0.056  &  0.016   & 0.286    & 0.050    & 0.286    & 0.050   & -1.96\\
 VIRMOS0.6GHz\_J022539-042548 &   0.057  &  0.015   & 0.203    & 0.039    & 0.203    & 0.039   & -1.53\\
 VIRMOS0.6GHz\_J022539-044528 &   0.067  &  0.017   & 0.281    & 0.047    & 0.281    & 0.047   & -1.73\\
 VIRMOS0.6GHz\_J022552-044233 &   0.068  &  0.017   & 0.265    & 0.045    & 0.265    & 0.045   & -1.64\\
 VIRMOS0.6GHz\_J022556-043522 &   0.056  &  0.018   & 0.235    & 0.041    & 0.235    & 0.041   & -1.73\\
 VIRMOS0.6GHz\_J022559-044421 &   0.078  &  0.020   & 0.274    & 0.051    & 0.274    & 0.051   & -1.51\\
 VIRMOS0.6GHz\_J022626-043657 &   0.069  &  0.016   & 0.211    & 0.041    & 0.211    & 0.041   & -1.35\\
 VIRMOS0.6GHz\_J022635-042049 &   0.066  &  0.017   & 0.266    & 0.044    & 0.266    & 0.044   & -1.68\\
 VIRMOS0.6GHz\_J022637-041336 &   0.070  &  0.018   & 0.329    & 0.046    & 0.329    & 0.046   & -1.86\\
 VIRMOS0.6GHz\_J022641-041602 &   0.073  &  0.016   & 0.296    & 0.046    & 0.296    & 0.046   & -1.69\\
 VIRMOS0.6GHz\_J022646-041156 &   0.064  &  0.016   & 0.295    & 0.051    & 0.295    & 0.051   & -1.84\\
 VIRMOS0.6GHz\_J022648-042040 &   0.064  &  0.016   & 0.240    & 0.045    & 0.240    & 0.045   & -1.59\\
 VIRMOS0.6GHz\_J022715-041648 &   0.061  &  0.016   & 0.283    & 0.053    & 0.283    & 0.053   & -1.85\\
\hline
\end{tabular}
\end{table*}
\section{Conclusions}

We observed the VVDS-VLA field (F02) with the
Giant Meter-Wave Radio Telescope at 610 MHz, imaging the 1 square degree area
with an angular resolution of 6 arcsec and an average sensitivity of about
50 $\mu$Jy/beam. A complete catalogue of 514 radio sources down to a local
$5\sigma$ limit has been compiled and the radio source counts at 610 MHz
have been derived. The source counts are the deepest obtained so far at this
frequency and clearly show the flattening of the radio counts at about 1 mJy.

The same field was previously observed with the VLA at 1.4 GHz with the same
angular resolution. We searched for 610 MHz counterparts of the 1054 radio
sources above $5\sigma$ in the 1.4 GHz catalogue, obtaining a spectral index
catalogue containing 741 matches
(448 above $5\sigma$ and 293 in the range 3-5$\sigma$ at 610 MHz) and 313
lower limits (below $3\sigma$ at 610 MHz).
Examining the spectral index properties of this catalogue as a function of the
radio flux density at 1.4 GHz and of the optical identification we found the 
following results:
\begin{itemize}
\item
Above 0.5 mJy the median spectral index is steep ($\alpha=-0.67\pm 0.05$) 
and is consistent with being associated with a population of classical radio 
loud AGNs.
\item
Between 0.15 mJy and 0.50 mJy the median radio spectral index becomes 
significantly flatter ($\alpha=-0.46\pm 0.03$). It is intriguing that the 
change in spectral index occurs at the same flux density corresponding to the
flattening of the radio source counts at 1.4 GHz, suggesting that the new
population of radio sources has different
spectral properties of the classical radio loud AGNs.
In particular, the flatter spectral index seems to rule out the starburst
galaxies as the dominant population of radio sources in this flux density
interval.
\item
At the faintest end of the flux density distribution (objects between 0.10
and 0.15 mJy), the median spectral index steepens again
($\alpha=-0.61\pm 0.04$). Unfortunately, incompleteness effects in this flux 
density interval do not allow us to draw any strong conclusions for these 
faintest
sources. Nevertheless, it is intriguing to suggest that the possible 
steepening of the spectral below 0.15 mJy might be associated with the
population of strong evolving late type and starburst objects.
\item
About 70\% of the radio sources are identified with optical galaxies divided
into three major categories: early-type, late-type, and starburst.
As expected, late-type and starburst galaxies have indistinguishable
spectral index distributions with a median spectral index $\alpha\simeq -0.70$,
while early-type galaxies have a flatter median spectral index 
$\alpha=-0.55\pm 0.04$. The unidentified objects have spectral properties
very similar to those of the radio sources identified with early-type galaxies.

\end{itemize}

Finally we constructed a sample of 58 candidate USS sources that need
further observations to be properly identified as real ultra-steep spectrum
sources. The most relevant feature of these candidate USS sources is
that they are between one and two orders of magnitude fainter than the USS 
sources used to find HzRGs. Therefore, they could be associated with even more
distant HzRGs, or with low luminosity HzRGs. In both cases these sources
represent almost unique objects and it will be very important to confirm
their nature of being USS sources and possible HzRGs.
In the meantime, we  show that there is a general trend for
these candidate USS sources to have fainter I-band and K-band magnitudes with
respect to the rest of the sources.    

\begin{acknowledgements}
We thank the staff of the GMRT who have made these observations possible. 
GMRT is run by the National Center for Radio Astrophysics of the Tata 
Institute of Fundamental Research.

%This work was performed under the framework of the VIRMOS consortium,
%and was supported by the Italian Ministry for University and Research
%(MURST) under grant COFIN-2000-02-34.

\end{acknowledgements}

\end{document}